\documentclass{article}

\usepackage{amssymb}

\begin{document}

\title{\bf Spherical Deconstruction}

\author{R.P. Andrews \footnote{pyrich@swan.ac.uk} \\ \\ Department of
  Physics, University of Wales Swansea \\ Singleton Park, 
Swansea, SA2 8PP, UK \\ \\ 
\and N. Dorey \footnote{N.Dorey@damtp.cam.ac.uk} \\ \\ Department of
Applied Mathematics and Theoretical Physics \\ 
Centre for Mathematical Sciences, University of Cambridge \\ Wilberforce Road, Cambridge, CB3 0WA, UK}
\date{}

\begin{figure}
\begin{flushright}
SWAT-05-432 \\
hep-th/0505107
\end{flushright}
\end{figure}

\maketitle

\begin{abstract}
We present evidence that ${\mathcal N}=1^{*}$ SUSY Yang-Mills provides a
deconstruction of a six-dimensional gauge
theory compactified on a two-sphere.  The six-dimensional theory 
is a twisted compactification of ${\mathcal N}=(1,1)$ SUSY Yang-Mills
theory of the type considered by Maldacena and Nunez (MN) in \cite{MN}.
In particular, we calculate the full classical spectrum of the 
${\mathcal N}=1^{*}$ theory with gauge group $U(N)$ in its Higgs vacuum.
In the limit $N \to \infty$, we find an exact agreement with the 
Kaluza-Klein spectrum of the MN compactification.   
\end{abstract}

\newpage

\section{Introduction}

One of the interesting consequences of recent developments in string
theory is the possibility that consistent (non-gravitational) quantum
theories exist in dimensions greater than four \cite{LST}. The idea of 
{\em deconstruction} \cite{AHCG} is a promising approach to defining
these theories
as certain limits of four-dimensional gauge theories\footnote{For
  other relevant work see \cite{Hal}.}. Typically, the
four-dimensional theory in its Higgs phase can be reinterpreted as 
a theory with additional compact, discretised directions. 
The hope is that higher-dimensional Lorentz invariance is recovered 
in an approporiate continuum limit.  
\paragraph{}
In this paper we will study a version of deconstruction which leads to
a six-dimensional gauge theory compactified to four-dimensions on 
a sphere. The starting point is $\mathcal{N}=1^{*}$ SUSY Yang-Mills
theory with gauge group $U(N)$. This theory contains a $U(N)$ gauge
multiplet together with three adjoint chiral multiplets of equal mass
$M$. Dependence of the superpotential on the mass parameter 
can be absorbed by a suitable rescaling of the fields. 
The superpotential then takes the form, 
\begin{equation}
W={\rm Tr}_{N}
\left(i\Phi_{1}[\Phi_{2},\Phi_{3}]+\frac{1}{2}\sum_{i=1}^{3}
\Phi_{i}^{2}\right)
\label{sp1}
\end{equation}    
The F-flatness condition is, 
\begin{equation} 
[\Phi_{i},\Phi_{j}]=i\varepsilon_{ijk}\Phi_{k}
\label{alg}
\end{equation}
As (\ref{alg}) coincides with the Lie algebra of $SU(2)$, it can be
solved by any $N$-dimensional representation of the $SU(2)$
generators \cite{VW}. This choice also solves the D-flatness condition. 
We will consider the vacuum corresponding to the solution 
$\Phi_{i}=J_{i}^{(N)}$. Here and in the following 
$J_{i}^{(n)}$ denote the generators of the
unique irreducible representation of $SU(2)$ of dimension 
$n$. In this ground state, the $U(N)$ gauge group is broken to 
the central $U(1)$ by the Higgs mechanism. We will refer to this state
as the Higgs vacuum.   
\paragraph{}
The emergence of additional dimensions in this model occurs in a way 
which is very familiar in the context of M(atrix) theory 
\cite{WATI,Kimura}, and was first interpreted in terms of 
deconstruction in \cite{AF}. 
Defining rescaled fields $\hat{x}_{i}=2\Phi_{i}/\sqrt{N^{2}-1}$, the 
expectation values in the Higgs vacuum satisfy, 
\begin{equation}
\hat{x}_{1}^{2}+\hat{x}_{2}^{2}+ \hat{x}_{3}^{2}={\bf 1}  
\label{ncs2}
\end{equation}
This is the defining equation of the fuzzy two-sphere \cite{Madore}. The
fuzzy sphere is a discrete, non-commutative version of the usual 
two-sphere. As we review below, expanding the $N \times N$ matrix 
fields of the ${\mathcal N}=1^{*}$ theory around this background 
naturally leads to a description in terms of fields defined on the 
fuzzy sphere. 
The resulting six-dimensional theory is non-commutative and has a UV
cutoff for finite $N$. Taking the limit $N\rightarrow \infty$ at the
classical level we obtain a commutative, continuum theory on 
$\Re^{3,1}\times S^{2}$. 
\paragraph{}
The appearance of a six-dimensional theory can also be understood 
via the string theory realisation of the ${\mathcal N}=1^{*}$ theory 
\cite{PS}. The theory is realised on the worldvolume of $N$ D3 branes    
in the presence of a background three-form field strength leading 
to a version of the Myers effect, where the D3 branes polarize into 
a spherically wrapped D5 brane \cite{Myers}. 
The theory living on the worldvolume 
of the D5 brane reduces to a six-dimensional $U(1)$ gauge
theory at low energies. The D3 brane charge is realised as $N$ units
of magnetic flux through the two-sphere leading to non-commutativity
in the world-volume gauge theory \cite{SW} as expected.  
\paragraph{} 
The main aim of this paper is to understand more precisely what kind of 
six-dimensional theory emerges from the set-up described above.  
The string theory picture suggests that 
we should start from a $U(1)$ gauge theory in six dimensions 
with ${\mathcal N}=(1,1)$ supersymmetry. This is the theory living 
on a single D5 brane with world-volume $\mathcal{R}^{5,1}$. 
This theory should then be compactified on 
a two-sphere in a way which preserves ${\mathcal N}=1$ supersymmetry 
in the four non-compact dimensions. A conventional
compactification would break all the supersymmetry 
as there is no covariantly constant spinor on $S^{2}$. 
On the other hand, a {\em twisted} compactification of the 
type which occurs when a D5-brane is wrapped on a noncontractible 
two-cycle of a Calabi-Yau threefold is known to preserve the 
required fraction of the supersymmetry \cite{BVS}. This 
compactification of the ${\mathcal N}=(1,1)$ theory, which 
was studied by Maldacena and Nunez (MN) in \cite{MN}, 
is therefore a natural candidate for the theory we seek. 
\paragraph{}
In this paper we will test this hypothesis 
by calculating the complete classical spectrum of the 
${\mathcal N}=1^{*}$ theory in the Higgs vacuum\footnote{For a related 
calculation in the context of SUSY quantum mechanics see \cite{porrati}
.} and comparing it with the spectrum of the twisted compactification 
of the six-dimensional theory described above. We find 
that the masses and degeneracies of states in 
the two theories agree exactly for $N=\infty$. This result suggests that 
the Higgsed ${\mathcal N}=1^{*}$ theory 
may be classically equivalent to a twisted $S^{2}$-compactification 
of a six-dimensional $U(1)$ gauge theory 
in this limit. The radius $R$ of the two-sphere of
is identified with $M^{-1}$, the inverse mass parameter of the
four-dimensional theory. This proposal could be tested further by explicitly 
comparing the Lagrangians of the two theories \cite{wip}.
Remarkably, the agreement of the two spectra also holds for any value 
of $N$ provided we truncate the expansion of the six-dimensional
fields in spherical harmonics appropriately. 
This truncation suggest that a six-dimensional description 
is valid for finite $N$ on length-scales larger than $L_{UV}\sim R/N$. 
\paragraph{}
Our discussion of deconstruction in this paper will be purely
classical. Of course the resulting six-dimensional gauge theory is
non-renormalisable and does not make sense at the quantum level
without some consistent UV completion. As mentioned above, the 
six-dimensional theory obtained at finite $N$ automatically has a
short-distance cut-off $L_{UV}$ of order $R/N$.  
To obtain a consistent quantum
theory we should seek a continuum limit where
physical scales of the six-dimensional theory are held fixed as
$L_{UV}\rightarrow 0$. 
In particular, we should hold the radius $R$ and the six-dimensional gauge 
coupling $G_{6}^{2}\sim g^{2}R^{2}/N$ fixed\footnote{
Here $g^{2}$ is the four-dimensional U(N) coupling. The low-energy 
effective gauge coupling of the unbroken $U(1)$ is $g^{2}/N$. The
latter is then identified with the effective four-dimensional $U(1)$ coupling 
$G^{2}_{4}=G_{6}^{2}/4\pi R^{2}$ arising from compactification of the
six dimensional theory leading to the estimate of $G_{6}^{2}$ given in
the text.}. This requires taking $N\rightarrow \infty$ with $g^{2}/N$
held fixed, which is a
strong-coupling limit of the four-dimensional theory. 
In \cite{nd2}, one of the
authors studied this strongly-coupled limit for a closely related model which
deconstructs a toroidal compactification of six-dimensional gauge
theory. Using S-duality and the AdS/CFT correspondence the proposed 
continuum limit was related to the decoupling limit of the
Neveu-Schwarz fivebrane which defines Little String Theory (LST). 
It is natural to conjecture that a
similar treatment of the ${\mathcal N}=1^{*}$ theory would lead to the
twisted spherical compactification of LST considered in \cite{MN}. 
In particular, this LST reduces at low-energies to the compactified 
six-dimensional gauge theory we discuss in this paper. The classical 
results described below provide some motivation for future work along
these lines.      
\paragraph{}
In the next two Sections 
we consider in turn the six-dimensional and four-dimensional
theories. Some additional information about vector spherical 
harmonics is relegated to an Appendix    
 
\section{Twisted Compactification}
In this Section we will describe the 
Maldacena-Nu$\tilde{\textrm{n}}$ez (MN) 
compactification of $\mathcal{N}=(1,1)$ SUSY Yang-Mills in six 
dimensions and find its classical spectrum of Kaluza-Klein modes. 
We start from the $U(1)$ theory defined on six-dimensional Minkowski 
space $\Re^{5,1}$. The global symmetry group is
\begin{equation}
SO(5,1) \times SO(4) \simeq SU(4) \times SU(2)_A \times SU(2)_B 
\end{equation}
Here $SO(5,1)$ is the six dimensional Lorentz group and 
$SO(4)$ is the R-symmetry of the ${\mathcal N}=(1,1)$ superalgebra. 
The matter content is a single $U(1)$ vector multiplet of 
$\mathcal{N} = (1,1)$ supersymmetry. It contains
\footnote{The corresponding spacetime indices run over
$M=0,1,\dots, 5$, $i=1,\dots,4$, $l=1,\dots,4$, $\bar{l}=1,\dots,4$} 
a six dimensional gauge
field $A_M$, four real scalar fields $\rho_{i}$ and two 
Weyl spinors of opposite chirality; $\lambda_{l}$ and 
$\tilde{\lambda}_{\bar{l}}$. 
Their transformation properties under the global symmetries are
\begin{center}
\begin{tabular}{l c c c}
\hline
& $SU(4)$ & $SU(2)_A$ & $SU(2)_B$ \\
\hline
$A_M$ & ${\bf 6}$ & ${\bf 1}$ & ${\bf 1}$ \\
$\rho_i$ & ${\bf 1}$ & ${\bf 2}$ & ${\bf 2}$ \\
$\lambda_l$ & ${\bf 4}$ & ${\bf 2}$ & ${\bf 1}$\\
$\tilde{\lambda}_{\bar{l}}$ & ${\bf \bar{4}}$ & ${\bf 1}$ & ${\bf 2}$\\
\hline
\end{tabular}
\end{center}
\paragraph{}
Anticipating compactification of two spatial dimensions, we write the
space-time as $\Re^{5,1} \sim \Re^{3,1} \times \Re^2$. 
This decomposition breaks the six-dimensional Lorentz group down to 
a subgroup, 
\begin{equation}
H=SO(3,1) \times SO(2) 
\end{equation}
with covering group, 
\begin{equation}
\bar{H}=SU(2)_L \times SU(2)_R \times U(1)_{45}
\end{equation}
It is straightforward to decompose 
the six-dimensional fields into representations of $H$. 
The gauge field is written as, 
\begin{eqnarray}
A_M & = & A_{\mu}  \qquad \mu = 0,1,2,3\\
\nonumber & = & A_a \qquad \ a = M-3 = 4,5
\end{eqnarray}
and we define the complex fields
\begin{equation}
n_{\pm} = \frac{1}{\sqrt{2}} \left( A_4 \pm i A_5 \right)
\end{equation}
Under the decomposition of the $SU(4)$ covering group,  
the six-dimensional spinors $\lambda_{l}$ and 
$\tilde{\lambda}_{\bar{l}}$, transforming in the 
${\bf 4}$ and $\bar{\bf 4}$, split according to, 
\begin{eqnarray}
{\bf 4}  & \rightarrow & ({\bf 2},{\bf 1})^{+1}\oplus 
({\bf 1},{\bf 2})^{-1} \nonumber \\ 
 \bar{\bf 4}  & \rightarrow & ({\bf 2},{\bf 1})^{-1}\oplus 
({\bf 1},{\bf 2})^{+1}  \nonumber
\end{eqnarray}
under $SU(2)_{L}\times SU(2)_{R}$ where the superscript denotes 
$U(1)_{45}$ charge. 
Thus we obtain a total of four left-handed Weyl spinors 
$\lambda^{A}_{\alpha}$,
$\psi^{\tilde{A}}_{\alpha}$, where $A=1,2$ and four right-handed
spinors $\bar{\lambda}^{A}_{\dot{\alpha}}$ and 
$\bar{\psi}^{\tilde{A}}_{\dot{\alpha}}$ with 
$\tilde{A}=1,2$. Here $A$ and $\tilde{A}$ are indices of 
$SU(2)_{L}$ and $SU(2)_{R}$ respectively.     
\paragraph{}
To summarize, the resulting bosonic fields then have quantum numbers, 
\begin{center}
\begin{tabular}{l c c c c c}
\hline
 & $SU(2)_L$ & $SU(2)_R$ & $U(1)_{45}$ & $SU(2)_A$ & $SU(2)_B$\\
\hline
$A_{\mu}$ & ${\bf 2}$ & ${\bf 2}$ & ${ 0}$ & ${\bf 1}$ & ${\bf 1}$ \\
$n_{\pm}$ & ${\bf 1}$ & ${\bf 1}$ & ${\pm 2}$ & ${\bf 1}$ & ${\bf 1}$ 
\\
$\rho_i$ & ${\bf 1}$ & ${\bf 1}$ & ${ 0}$ & ${\bf 2}$ & ${\bf 2}$ \\
\hline
\end{tabular}
\end{center}
while the fermions transform as, 
\begin{center}
\begin{tabular}{l c c c c c}
\hline
 & $SU(2)_L$ & $SU(2)_R$ & $U(1)_{45}$ & $SU(2)_A$ & $SU(2)_B$\\
\hline
$\lambda^A_{\alpha}$ & ${\bf 2}$ & ${\bf 1}$ & ${ +1}$ & ${\bf 2}$ & 
${\bf 1}$ \\
$\bar{\lambda}^A_{\dot{\alpha}}$ & ${\bf 1}$ & ${\bf 2}$ & ${ -1}$ & 
${\bf 2}$ & ${\bf 1}$ \\
$\psi^{\tilde{A}}_{\alpha}$ & ${\bf 2}$ & ${\bf 1}$ & ${ -1}$ & ${\bf 1}$
 & ${\bf 2}$ \\
$\bar{\psi}^{\tilde{A}}_{\dot{\alpha}}$ & ${\bf 1}$ & ${\bf 2}$ & ${+1}$
 & ${\bf 1}$ & ${\bf 2}$ \\
\hline
\end{tabular}
\end{center}
\paragraph{}
We now want to compactify the theory by replacing the $45$ plane 
by a two-sphere. In conventional compactification the couplings of
fields to the curvature of the sphere are determined by their 
quantum numbers under $U(1)_{45}$ which corresponds to local rotations
in the $45$ plane. In the compactification of Maldacena and 
Nu$\tilde{\textrm{n}}$ez the theory is twisted by embedding the local
rotation group into the $SU(2)_{A}\times SU(2)_{B}$ R-symmetry group
of the theory. To accomplish this we define 
Cartan subgroups $U(1)_A$ and $U(1)_{B}$ 
of $SU(2)_A$ and $SU(2)_{B}$ with corresponding generators $Q_{A}$ and
$Q_{B}$ respectively\footnote{These generators are normalised to take
  the values $Q_{A}=\pm 1$ on states in the fundamental representation
  of $SU(2)$.}. We also define the diagonal subgroup 
$U(1)_T = D(U(1)_{45} \times U(1)_A)$ with generator $Q_T = Q_{45} +
Q_A$. The vector multiplet fields then have quantum numbers,  
\begin{center}
\begin{tabular}{l c c}
\hline
  & $U(1)_A$ & $U(1)_T$ \\
\hline
$A_{\mu}$ & $0$ & ${ 0}$  \\
$n_{\pm}$ & $0$ & ${ \pm 2}$  \\
$\rho_i$ & $\pm 1$ & ${\pm 1}$ \\
$\lambda^A_{\alpha}$ & $\pm 1$ & $
\left( \begin{array}{c} { +2} \\ { 0} \end{array}\right)$ \\
$\bar{\lambda}^A_{\dot{\alpha}}$ & $\pm 1$ & $\left( 
\begin{array}{c} {0} \\ { -2} \end{array}\right)$ \\
$\psi^{\tilde{A}}_{\alpha}$ & $0$ & ${ -1}$ \\
$\bar{\psi}^{\tilde{A}}_{\dot{\alpha}}$ & $0$ & ${ +1}$ \\
\hline
\end{tabular}
\end{center}
\paragraph{}
We then compactify the theory with $U(1)_{T}$ playing the role of the 
local rotation group (instead of $U(1)_{45}$). We will refer to the
$U(1)_{T}$ quantum number as T-spin and the six-dimensional fields 
can be split up accordingly as,   
\begin{center}
\begin{tabular}{l l c l}
T-scalars: & $Q_T = 0$ && $A_{\mu}$, $\lambda^{A=2}_{\alpha}$, 
$\bar{\lambda}^{A=1}_{\dot{\alpha}}$\\
T-spinors: & $Q_T = \pm 1$ && $\psi^{\tilde{A}}_{\alpha}$, 
$\bar{\psi}^{\tilde{A}}_{\dot{\alpha}}$, $\rho_i$ \\
T-vectors: & $Q_T = \pm 2$ && $n_{\pm}$, $\lambda^{A=1}_{\alpha}$, 
$\bar{\lambda}^{A=2}_{\dot{\alpha}}$
\end{tabular}
\end{center}
Correspondingly the terms scalar, spinor and vector 
will be reserved for describing
the transformation properties of fields under the four-dimensional 
Lorentz group. The existence of a single Weyl spinor (of both
chiralities) which is also a T-scalar guarentees the existence of a 
single massless fermion in four-dimensions as required by 
${\mathcal N}=1$ supersymmetry. Each six dimensional field has a 
kinetic term on $S^{2}$. After expanding in appropriate spherical 
harmonics, this kinetic term determines the masses of an infinite tower
of four-dimensional fields. 
We now consider the Kaluza-Klein spectrum of each type
of field in turn.  
\paragraph{}
After integration by parts, the kinetic term for a T-scalar field 
$\varphi$ defined on a two-sphere of unit radius can be written as,  
\begin{equation}
S_{\varphi}= \int \, d^{2}\Omega \,\, \varphi \Delta_{S^{2}} \varphi 
\end{equation}
where  $\Delta_{S^{2}}$ is the scalar Laplacian on the two-sphere. 
To find the mass eigenstates we expand $\varphi$ 
in terms of spherical harmonics as, 
\begin{eqnarray}
 \varphi(\theta,\phi) & = & 
\sum^{\infty}_{l=0} \sum^{+l}_{m=-l} \varphi_{lm} Y_{lm}(\theta,\phi)\\
\nonumber 
\end{eqnarray}
where $(\theta,\phi)$ are the standard spherical polar coordinates.  
$Y_{lm}$ is an eigenfunction of the Laplacian $\Delta_{S^2}$ with
eigenvalue $l(l+1)$ for integer $l\geq 0$ and for $m=-l,\ldots +l$. 
Thus for each T-scalar field in six dimensions we find a 
Kaluza-Klein tower of four dimensional 
fields with masses\footnote{The masses are dimensionless as we have
  set the radius of the two-sphere to be unity.} 
$l(l+1)$ with degeneracy $(2l+1)$. According to the 
list of T-scalar fields given above we find a four-dimensional vector
field, a left-handed Weyl spinor and a right-handed Weyl spinor at each 
mass level. The corresponding representations of the four-dimensional 
Lorentz group $SU(2)_{L}\times SU(2)_{R}$ are,
\begin{eqnarray*}
(\mathbf{2},\mathbf{2}) \oplus (\mathbf{2},\mathbf{1}) 
\oplus (\mathbf{1},\mathbf{2})
\end{eqnarray*} 
For $l=0$, the corresponding four-dimensional fields are massless 
and the spin quantum numbers match those of 
a single massless vector multiplet of ${\mathcal N}=1$ supersymmetry.   
For $l>1$, we find massive vector fields together with Weyl 
fermions. However, a massive vector multiplet of ${\mathcal N}=1$
supersymmetry also includes scalar fields in four-dimensions. 
Thus, for $l>0$, the fields descending from the T-scalars in six 
dimensions do not form complete multiplets of ${\mathcal N}=1$ 
supersymmetry. This puzzle will be resolved below where we will find 
the additional states needed to form massive vector multiplets.    
\paragraph{}
A two-component Dirac T-spinor $\Psi$ defined on a two-sphere has
kinetic term, 
\begin{equation}
S_{\Psi}=\int\, d^{2}\omega\,\, i \bar{\Psi}\hat{\nabla}_{S^2} \Psi 
\label{dirac2}
\end{equation}
The Dirac operator on the sphere is defined as, 
\begin{equation}
i \hat{\nabla}_{S^2} = i e^{\mu a} \gamma_a \nabla_{\mu}
\end{equation}
where $e^{\mu a}$ is the zweibein, $\gamma_{a}$ represent the Clifford
algebra and the spinor covariant derivative on $S^{2}$ is given as, 
\begin{equation}
\nabla_{\mu} \Psi = \partial_{\mu} \Psi + \frac{i}{4} R^{ab}_{\mu} 
\sigma_{ab} \Psi
\end{equation}
where $R_{ab}$ is the spin-connection and $\sigma_{ab}$ are the
rotation generators. An explicit expression for the Dirac operator in
spherical polars is given in \cite{Abri} as,  
\begin{equation}
i \hat{\nabla}_{S^2} = i \sigma_1 \left(\frac{\partial}
{\partial \theta} + \frac{\textrm{cot} \theta}{2} \right) 
+ \frac{i \sigma_2}{\textrm{sin} \theta} \frac{\partial}{\partial \phi}
\end{equation}
where $\sigma_{1}$ and $\sigma_{2}$ are Pauli matrices. 
A Dirac T-spinor field on $S^{2}$ can be expanded in terms of a 
complete basis formed by the eigenfunctions of the squared Dirac 
operator,   
\begin{equation} 
\left(-i \hat{\nabla}_{S^2}\right)^{2}\Psi=\mu \Psi 
\end{equation}
The allowed eigenvalues correspond to $\mu=l^{2}$ for 
integer $l\geq 1$ with degeneracy $2l$ \cite{Abri}. 
\paragraph{}
The fermionic T-spinor fields, $\psi^{\tilde{A}}_{\alpha}$ and 
$\bar{\psi}^{\tilde{A}}_{\dot{\alpha}}$ listed above can be 
combined to form two-component Dirac spinors on $S^{2}$ with kinetic
terms (\ref{dirac2}) according to, 
\begin{equation}
\rho^{(\tilde{A})}_{\alpha}=\left(\begin{array}{l} 
\psi^{\tilde{A}}_{\alpha} \\
\bar{\psi}^{\tilde{A}}_{\dot{\alpha}=\alpha}
\end{array}\right)
\end{equation}
for $\tilde{A}=1,2$, $\alpha=1,2$. Thus we obtain four-species of Dirac
spinors on the sphere. Each species yields $2l$ states of mass 
$\mu=l^{2}$ for $l\geq 1$ after expansion in terms of eigenstates of
the squared Dirac operator. At each mass level we therefore find
$8l$ off-shell degrees of freedom which can be recombined as $4l$
left-handed and $4l$ right-handed Weyl spinors in four dimensions. 
These Weyl spinors must 
be paired with bosonic fields to form multiplets of 
${\mathcal N}=1$ SUSY in four dimensions. 
The extra fields come from Kaluza-Klein reduction of the bosonic 
T-spinors $\rho_{i}$, $i=1,2,3,4$, which yield massive scalar fields in 
four dimensions. These states combine with the fermionic T-spinors to 
form massive chiral multiplets with Lorentz spins,  
\begin{eqnarray*}
(\mathbf{2},\mathbf{1}) \oplus (\mathbf{1},\mathbf{2}) \oplus 
2 \times (\mathbf{1},\mathbf{1})
\end{eqnarray*}
\paragraph{}
It remains to determine the Kaluza-Klein spectrum of the T-vector
fields. As we have unbroken ${\mathcal N}=1$ supersymmetry in the
four non-compact dimensions it suffices to focus on the 
bosonic T-vector fields $n_{\pm}=(A_{4}\pm iA_{5})/\sqrt{2}$. 
The two real components $A_{4}$ and $A_{5}$ define a Maxwell gauge 
field $a_{\rho}$ on the two-sphere. The resulting 
kinetic term reads, 
\begin{equation}
S_{A} = \int d^{2} \Omega\,\,\frac{1}{4} F_{\rho\kappa} F^{\rho\kappa}
\end{equation}
where $\rho,\kappa=1,2$ label coordinates on 
$S^{2}$ and $F_{\rho\kappa}=
\partial_{\rho}a_{\kappa}-\partial_{\kappa}a_{\rho}$. 
We impose the gauge condition $\nabla^\rho a_\rho = 0$ and expand
${\bf a}=(a_{1},a_{2})$ in terms of vector spherical harmonics,   
\begin{eqnarray}
\nonumber {\bf a} & = & \sum_{l,m} a_{lm} {\bf T_{lm}} 
\end{eqnarray}
The gauge-fixed spherical harmonics 
${\bf T}_{lm}$ appearing in this expansion are defined
in Appendix A. This expansion diagonalises the Maxwell term as,  
\begin{eqnarray}
S_{A} = \sum_{l,m,l',m'} 
a^{\dag}_{lm} a^{\phantom{\dag}}_{l'm'} l(l+1) \delta_{ll'} \delta_{mm'}
\end{eqnarray}
Thus the T-vector field $n_{\pm}$ yields a Kaluza-Klein tower of 
four-dimensional scalar fields of mass $l(l+1)$ with degeneracy $2l+1$
for $l\geq 1$. Notice that these fields are degenerate in mass with
the four-dimensional vector fields coming from the Kaluza-Klein
reduction of the T-scalars. In fact the number of scalar fields is
just right to pair up with the massive vector fields to form massive 
vector multiplets of ${\mathcal N}=1$ SUSY in four dimensions with 
Lorentz spins,  
\begin{eqnarray*}
(\mathbf{2},\mathbf{2}) \oplus 2 \times [(\mathbf{2},\mathbf{1}) 
\oplus (\mathbf{1},\mathbf{2})] \oplus (\mathbf{1},\mathbf{1})
\end{eqnarray*}
The fermionic part of each multiplet includes two species of left- and
right-handed Weyl fermions. One species comes from the KK reduction
of the fermionic T-scalars  $\lambda^{A=2}_{\alpha}$ and 
$\bar{\lambda}^{A=1}_{\dot{\alpha}}$ and the other comes from the
reduction of the fermionic T-vectors $\lambda^{A=1}_{\alpha}$, 
$\bar{\lambda}^{A=2}_{\dot{\alpha}}$. 
\paragraph{}
We summarise the complete Kaluza-Klein spectrum of the MN
compactification in the table below, 
\begin{center}
\end{center}
T-scalar:
\begin{center}
\begin{tabular}{c|c}
$\lambda$ & States \\
\hline
$l(l+1)$ & $(2l+1) \times$ $\Big\lbrace ({\bf 2},{\bf 2}) \oplus 
({\bf 2},{\bf 1}) \oplus ({\bf 1}, {\bf2}) \Big\rbrace$
\end{tabular}\\
$l = 0,1,2,\dots$ \qquad
\end{center}
T-spinor:
\begin{center}
\begin{tabular}{c|c}
$\lambda$ & States \\
\hline
$l^2$ & $4l \times$ $\Big\lbrace ({\bf 2},{\bf 1}) \oplus ({\bf 1},
{\bf 2}) \oplus 2 \times ({\bf 1},{\bf 1}) \Big\rbrace$
\end{tabular}\\
$l = 1,2,3,\dots$
\end{center}
T-vector:
\begin{center}
\begin{tabular}{c|c}
$\lambda$ & States \\
\hline
$l(l+1)$ & $(2l+1) \times$ $\Big\lbrace ({\bf 2},{\bf 1}) \oplus 
({\bf 1},{\bf 2}) \oplus 2 \times ({\bf 1},{\bf 1}) \Big\rbrace$
\end{tabular}\\
$l = 1,2,3,\dots$
\end{center}
\paragraph{}
We can also present the spectrum in terms of complete ${\mathcal N}=1$
multiplets. The spectrum includes a single massless $U(1)$ vector
multiplet. The massive spectrum is labelled by a positive integer
$l=1,2,\ldots$ and includes the following states,  
\begin{center}
\begin{tabular}{c c c}
\hline
Mass & Degeneracy & Multiplet \\
\hline
$l(l+1)$ & $2l+1$ & massive vector \\
$l^{2}$   & $4l$  & massive chiral \\
\hline
\end{tabular}
\end{center}

\section{The Spectrum $\mathcal{N} = 1^*$ SUSY Yang-Mills}
\paragraph{}
In this section we will compute the classical spectrum of the 
${\mathcal  N}=1^{*}$ theory in the Higgs vacuum described above. 
As the theory has ${\mathcal N}=1$ supersymmetry it suffices to
consider the mass matrix of the fermions in the theory. The relevant 
terms in the Lagrangian are, 
\begin{eqnarray}
\mathcal{L}_{F} & = & \textrm{Tr}_N \Big\lbrace i\psi_i \epsilon_{ijk}
[\Phi_k , \psi_j] + i \bar{\psi}_i \epsilon_{ijk}[\Phi^{\dag}_k , 
\bar{\psi}_j] - i\lambda [ \Phi^{\dag}_i , \psi_i ] \\ 
\nonumber & & + i\psi_i [\Phi^{\dag}_i , \lambda ] + i\bar{\psi}_i 
[ \Phi_i , \bar{\lambda} ] - i\bar{\lambda} [ \Phi_i , \bar{\psi}_i ]
 - \psi_i \psi_i - \bar{\psi}_i \bar{\psi}_i
\Big\rbrace
\end{eqnarray}
where $\lambda$ is the gaugino and $\psi_i$ are the superpartners of 
$\Phi_i$. In the Higgs vacuum, the scalar fields take expectation 
values $\Phi_{i}=J_{i}^{(N)}$ for $i=1,2,3$. The resulting mass terms 
for the fermions are written as,   
\begin{equation}
\mathcal{L}_{F} = \Big\lbrace \big(\Psi_R \big)_{ab} 
\Delta^{(R S)}_{ab,cd} \big(\Psi^T_S \big)_{cd} + \big(\bar{\Psi}_R 
\big)_{ab} \bar{\Delta}^{(R S)}_{ab,cd} \big(\bar{\Psi}^T_S \big)_{cd}
\Big\rbrace
\end{equation}
where the four species of Weyl fermion are combined in a column
$\Psi_{R}$ with $\Psi_R = \psi_i$ for $R=i=1,2,3$ and $\Psi_4 =
\lambda$. 
\paragraph{}
Our aim is to find a suitable basis in which to diagonalise the fermion
mass matrices $\Delta$ and $\bar{\Delta}$ defined above. 
The string theory picture of 
the Higgs vacuum described above suggests that the mass eigenstates 
should correspond to spherical harmonics on the two-sphere. As we now 
review, there is a natural map between $N\times N$
matrices and functions on the two-sphere which will provide the basis
we seek. Scalar functions on the two sphere can be expanded as,  
\begin{eqnarray}
a(\Omega) & = & \sum^{\infty}_{l=0} \sum^l_{m=-l} a_{lm}
Y_{lm}(\Omega) 
\nonumber
\end{eqnarray}
The spherical harmonics can be expressed in terms of the cartesian
coordinates $x_{A}$ with $A=1,2,3$ of a unit vector in $\Re^{3}$,  
\begin{eqnarray}
Y_{lm}(\Omega) & = & \sum_{\vec{A}} 
f ^{(lm)} _{A_1 \dots A_l} x^{A_1} \! \dots x^{A_l}
\end{eqnarray}
where $f ^{(lm)} _{A_1 \dots A_l}$ is a traceless symmetric tensor of 
$SO(3)$ with rank $l$. Similarly, 
$N\times N$ matrices can be expanded as,  
\begin{eqnarray}
\hat{a} & = & \sum^{N-1}_{l=0} \sum^l_{m=-l} a_{lm} \hat{Y}_{lm} \\
\hat{Y}_{lm} &
 = & \sum_{\vec{A}} f ^{(lm)} _{A_1 \dots A_l} 
\hat{x}^{A_1} \! \dots \hat{x}^{A_l}
\end{eqnarray}
where $\hat{x}_{A}=2J^{(N)}_{A}/\sqrt{N^{2}-1}$. 
The matrices $\hat{Y}_{lm}$ are known as 
fuzzy spherical harmonics and they obey the orthonormality condition,  
\begin{equation}
\textrm{Tr}_N \left( Y^{\dag}_{lm} Y_{l'm'}^{\phantom{\dag}} \right) = 
\delta_{l \, l'} \, \delta_{m \, m'}
\end{equation}
The coefficient $f ^{(lm)} _{A_1 \dots A_l}$ is the same as on the 
sphere. In this way we obtain a map from $N\times N$ 
matrices to functions on the two-sphere;  
\begin{equation}
\hat{a} = \sum^{N-1}_{l=0} \sum^{+l}_{m=-l} a_{lm} \hat{Y}_{lm} \to
a(\Omega) = 
\sum^{N-1}_{l = 0} \sum^l_{m = -l} a_{lm} Y_{lm}
\end{equation}
Notice that the expansion in spherical harmonics is truncated at 
$N-1$ reflecting the finite number of degrees of freedom in the matrix 
$\hat{a}$. 
\paragraph{}
Motivated by the above discussion we expand the fermionic fields 
$\Psi_R$ in fuzzy spherical harmonics as, 
\begin{equation}
\Psi_R = \sum_{lm} \hat{\Psi}^{(R)}_{lm} \hat{Y}_{lm}
\label{exp}
\end{equation}
The masses of physical states are detemined by squared mass matrix, 
\begin{equation}
M^{(R S)}_{ab, \, ef} = 
\bar{\Delta}^{(R T)}_{ab, \, cd} \Delta^{(T S)}_{cd, \, ef}
\end{equation}
In order to diagonalize this matrix we consider the bilinear form, 
\begin{equation}
{\mathcal B}=\big(\Psi^{\dag}_{R}\big)_{ab} M^{(R S)}_{ab, \, ef}
\big(\Psi^T_{S}\big)_{ef}
\end{equation}
The expansion (\ref{exp}) then yields, 
\begin{equation}
{\mathcal B}=
\sum^{N-1}_{l=0} \sum^{l}_{m=-l} \sum^{N-1}_{l'=0} \sum^{l'}_{m'=-l'} 
\big(\hat{\Psi}^{\, (R)}_{lm}\big)^{\dag} \hat{\Psi}^{\, (S)}_{l'm'} 
N^{(R S)}_{lm, \, l'm'}
\end{equation}
with\footnote{As above $J^{(L)}_i$ are the generators of $SU(2)$ in the 
irreducible representation of dimension $L$}
\begin{equation}
N^{(R S)}_{lm, \, l'm'}= \delta_{ll'}
\left( \begin{array}{cccc}
J_{(L)}^{\, 2} + 1 & -iJ^{(L)}_3 & iJ^{(L)}_2 & 0 \\
iJ^{(L)}_3 & J_{(L)}^{\, 2} + 1 & -iJ^{(L)}_1 & 0 \\
-iJ^{(L)}_2 & iJ^{(L)}_1 & J_{(L)}^{\, 2} +1 & 0 \\
0 & 0 & 0 & J_{(L)}^{\, 2}
\end{array} \right)_{mm'}
\end{equation}
with $L=2l+1$. 
\paragraph{}
To complete the diagonalization,
 we will determine the characteristic equation of the matrix 
$N^{(R S)}_{lm, \, l'm'}$. 
Consider the $(p+q)\times(p+q)$ matrix
\begin{equation}
{\mathcal X} = \begin{array}{cc}
& \begin{array}{cc} (p) & (q) \end{array} \\
\begin{array}{c} (p) \\ (q) \end{array} &
\left(\begin{array}{cc} 
A & B \\ C & D 
\end{array} \right)
\end{array}
\end{equation}
The determinant of ${\mathcal X}$ can be evaluated using the formula 
\cite{Part}
\begin{equation}
Det({\mathcal X}) = Det(D)Det(A-BD^{-1}C) = Det(A)Det(D-CA^{-1}B)
\end{equation}
Applying this formula in the present case we find, 
\begin{equation}
Det(N-\lambda \mathbf{1}) = \prod^{N-1}_{l=0} 
\prod^l_{m=-l} (\gamma^{(l)}-1)^{2(2l+1)} (\gamma^{(l)}+l)^{2l+3} 
\big(\gamma^{(l)}-(l+1)\big)^{2l-1} = 0
\end{equation}
where $\gamma^{(l)} = l(l+1) + 1 - \lambda$. 
The roots of this equation yield the eigenvalues of the fermion mass
matrix. For $l=0$ we therefore find, 
\begin{center}
\begin{tabular}{c c}
\hline
Eigenvalue & Degeneracy \\
\hline
$0$   & $1$ \\
$1$ & $3$ \\
\hline
\end{tabular}
\end{center}
while for $l=1,2,\ldots,N-1$ we get, 
\begin{center}
\begin{tabular}{c c}
\hline
Eigenvalue & Degeneracy \\
\hline
$l^{2}$   & $2l-1$ \\
$l(l+1)$ & $2(2l+1)$ \\
$(l+1)^{2}$   & $2l+3$ \\
\hline
\end{tabular}
\end{center}
To find the complete spectrum in this case 
we sum over all values of $l$. The final result is a single zero 
eigenvalue and two series 
of eigenvalues labeled by a positive integer $k=1,2,\ldots N-1$, 
\begin{center}
\begin{tabular}{c c}
\hline
Eigenvalue & Degeneracy \\
\hline
$k^{2}$   & $4k$ \\
$k(k+1)$ & $2(2k+1)$ \\
\hline
\end{tabular}
\end{center}
Finally we find one extra eigenvalue $\lambda=N^{2}$ with 
multiplicity $2N+1$. 
\paragraph{}
Each eigenvalue corresponds to a
single left-handed Weyl fermion and its right-handed charge
conjugate. As the theory has unbroken ${\mathcal N}=1$ supersymmetry we
must combine these states with bosons to form complete multiplets. 
The full bosonic spectrum can be calculated by similar means, but it
suffices to note the following constraints. In the vacuum state we are
considering, the $U(N)$ gauge group is broken down to $U(1)$. The 
spectrum therefore includes a single massless gauge boson and
$N^{2}-1$ massive gauge bosons. The Weyl fermions must combine with
these states to form ${\mathcal N}=1$ vector multiplets. Any left over
fermions must be paired with scalar fields to form chiral multiplets. 
\paragraph{}
Given the above
spectrum of fermions the only possibility consistent with ${\mathcal N}=1$ 
SUSY is as follows. In addition to a single massless vector multiplet 
we have two towers of multiplets labelled by $k=1,2,\ldots, N-1$ as
tabulated below,
\begin{center}  
\begin{tabular}{c c c}
\hline
Mass & Degeneracy & Multiplet \\
\hline
$k(k+1)$ & $2k+1$ & massive vector \\
$k^{2}$   & $4k$  & massive chiral \\
\hline
\end{tabular}
\end{center} 
The spectrum is completed by $2N+1$ chiral multiplets of mass 
$N^{2}$. In the limit $N\rightarrow \infty$ this precisely matches the 
spectrum of the Maldacena-Nunez compactification. For finite $N$ 
the spectrum of the ${\mathcal N}=1^{*}$ theory is a subset of that of 
the six-dimensional theory obtained by retaining only those states with
mass less than $N^{2}$ (together with $2N+1$ chiral multiplets of mass
equal to $N^{2}$).   
\paragraph{}
The work of ND is supported by a PPARC senior fellowship. 

\appendix

\section{Vector Harmonics}

We present a brief review of vector harmonics. The eigenfunctions of
the scalar Laplace operator on the sphere correspond to ordinary 
spherical harmonics $Y_{lm}$. To consider vector fields on the sphere
it is not consistent to expanded the components of the vector
seperately in scalar spherical harmonics. Instead a vector field
defined on $\Re^{3}$  can be expanded in 
terms of the following vector harmonics \cite{Mag,Astro}.
\begin{equation}
\tilde{{\bf a}}(\Omega) = \sum_{lm} \left( t_{lm} \tilde{{\bf T}}_{lm} 
+ s_{lm} \tilde{{\bf S}}_{lm} + r_{lm} \tilde{{\bf R}}_{lm} \right)
\end{equation}
where
\begin{eqnarray}
\tilde{{\bf R}}_{lm} & = & \hat{\bf r} Y_{lm} \\
\tilde{{\bf T}}_{lm} & = & \frac{1}{\sqrt{l(l+1)}} 
\left[-\frac{\partial Y_{lm}}{\partial \theta} 
\mbox{\boldmath $\hat{\phi}$} + \textrm{csc} \, \theta 
\frac{\partial Y_{lm}}{\partial \phi} 
\mbox{\boldmath $\hat{\theta}$} \right] \\
\tilde{{\bf S}}_{lm} & = & \frac{1}{\sqrt{l(l+1)}} 
\left[\frac{\partial Y_{lm}}{\partial \theta} 
\mbox{\boldmath $\hat{\theta}$} + \textrm{csc} \, \theta 
\frac{\partial Y_{lm}}{\partial \phi} \mbox{\boldmath $\hat{\phi}$}
\right]
\end{eqnarray}
Restricting to the two-sphere $r=1$ we find $\tilde{{\bf R}}_{lm} =
0$. Covariant and contravariant vectors on the two-sphere can then be
defined as, 
\begin{equation}
\tilde{V}_i = h_i V^i = h^{-1}_i V_i
\end{equation}
where $g_{ij} = h^2_i \delta_{ij}$. The corresponding 
covariant vector harmonics are
\begin{eqnarray}
{\bf T}_{lm} & = & \frac{1}{\sqrt{l(l+1)}} \left[-\textrm{sin} \, 
\theta \partial_{\theta} Y_{lm} \mbox{\boldmath $\hat{\phi}$} 
+ \textrm{csc} \, \theta \partial_{\phi} Y_{lm} 
\mbox{\boldmath $\hat{\theta}$} \right] \\
{\bf S}_{lm} & = & \frac{1}{\sqrt{l(l+1)}} \left[\partial_{\theta}
 Y_{lm} \mbox{\boldmath $\hat{\theta}$} + \partial_{\phi} Y_{lm} 
 \mbox{\boldmath $\hat{\phi}$} \right]
\end{eqnarray}
The Maxwell field on a two-sphere is a vector field with the gauge
invariance, 
\begin{equation}
A_{\mu} \to A_{\mu} - \partial_{\mu} \chi
\end{equation}
The gauge fields can be expanded in vector harmonics and the scalar
$\chi$ can be expanded in spherical harmonics. Under a gauge 
transformation the components transform as, 
\begin{eqnarray}
A'_{\theta}(\theta, \phi) & = & \sum_{lm} \left( t_{lm} \textrm{csc}
\, \theta \, \partial_{\phi} Y_{lm} + s_{lm} \partial_{\theta} Y_{lm}
- \chi_{lm} \partial_{\theta} Y_{lm} \right)  \nonumber \\
A'_{\phi}(\theta, \phi) & = & \sum_{lm} \left( t_{lm} (-\textrm{sin}
\, \theta) \partial_{\theta} Y_{lm} + s_{lm} \partial_{\phi} Y_{lm} -
\chi_{lm} \partial_{\phi} Y_{lm} \right) \nonumber
\end{eqnarray}
It therefore follows that we can set $s_{lm} = 0$ via a gauge
transformation with $\chi_{lm}=s_{lm}$. The corresponding 
gauge fixing condition is the generally covariant analogue of the 
Lorentz Gauge.
\begin{eqnarray}
\nonumber \nabla^a A_a & = & g^{ab} \nabla_b A_a \\
& = & g^{ab} \partial_b A_a - g^{ab} \Gamma^c_{ba} A_c
\end{eqnarray}
For the sphere there are only three non-zero Christoffel symbols
\begin{equation}
\Gamma^{\theta}_{\phi \phi} = - \textrm{cos} \, \theta \, \textrm{sin}
 \, \theta \qquad \Gamma^{\phi}_{\theta \phi} = 
 \Gamma^{\phi}_{\phi \theta} = \textrm{cot} \, \theta
\end{equation}
Therefore
\begin{eqnarray}
\nonumber \nabla^a A_a & = & g^{ab} \partial_b A_a + \textrm{cot} \, 
\theta A_{\theta} \\
\nonumber & = & \sum_{lm} \frac{1}{\sqrt{l(l+1)}} \Big\lbrace t_{lm} 
\Big( \partial_{\theta} (\textrm{csc} \, \theta \partial_{\phi} Y_{lm})
 + \textrm{cot} \, \theta \textrm{csc} \, \theta \partial_{\phi} Y_{lm}
 \\
\nonumber & & \ \ - \textrm{csc} \, \theta \partial_{\phi} 
\partial_{\theta} Y_{lm} \Big)  + s_{lm} \Big( \partial_{\theta} 
\partial_{\theta} Y_{lm} + \textrm{cot} \, \theta \partial_{\theta} 
Y_{lm} + \textrm{csc}^2 \theta \partial_{\phi} \partial_{\phi} Y_{lm}
 \Big) \Big\rbrace \\
& = & \sum_{lm} \frac{1}{\sqrt{l(l+1)}} \, s_{lm} \Delta_{S^2} Y_{lm}
\end{eqnarray}
The gauge condition $\nabla^a A_a = 0$ thus sets $s_{lm} = 0$.

\end{document}